\documentstyle[graphicx]{elsart}
\begin{document}
\begin{frontmatter}
\title{Torsional Oscillator Studies of the Superfluidity of $^3$He in
Aerogel}
\author[address1]{H Alles},
\author[address1]{J J Kaplinsky}
\author[address1]{P S Wootton}
\author[address1]{J D Reppy\thanksref{thank1}} \and
\author[address1]{J R Hook}
\address[address1]{Schuster Laboratory, University of Manchester,
Manchester, M13 9PL, U.K.}
\thanks[thank1]{present address:  LASSP, Clark Hall, Cornell
University, Ithaca, NY 14853, USA.}
\begin{abstract}

We have made simultaneous torsional oscillator and transverse NMR
measurements (at $\sim165\,{\rm kHz}$) on $^3$He contained within
aerogels with nominal densities of 1\% and 2\% of solid glass. The
superfluid transition is seen simultaneously by both techniques
and occurs at a temperature which agrees semi-quantitatively with
that expected for homogeneous isotropic pair-breaking scattering
of $^3$He atoms by strands of silica.
Values obtained for the superfluid density $\rho_{\rm s}$ in the 2\%
sample are in reasonable agreement with those observed previously.
Coupling of the torsional mode to a parasitic resonance prevented
accurate determination of $\rho_{\rm s}$ for the 1\% aerogel. We have
identified other resonances coupling to the torsional oscillations as
sound modes within the helium/aerogel medium.

\end{abstract}
\end{frontmatter}

\section{Introduction}
The discovery of superfluidity of $^3$He in low density aerogel glass
in torsional oscillator experiments at Cornell
University\cite{Porto95} and its subsequent investigation in NMR
experiments at Northwestern University\cite{Halperin95} raised many
interesting questions; for example: 

\begin{itemize}
\item Would the transition seen in the torsional oscillator be
coincident with the onset of an NMR frequency shift? 
\item How many superfluid phases are there
and what is the phase diagram?
\item What is the nature of the pairing?
\item Is it possible to explain theoretically the effect of the
aerogel on the properties of the superfluid phase(s)?
\end{itemize}
In an attempt to address these questions we decided to perform
simultaneous torsional oscillator and NMR experiments on the same
sample.

In this paper we report the results of torsional oscillator
measurements on aerogel samples with 1\% and 2\% of solid density.
Analysis of the NMR data is still in progress so we do not report
these results in detail here. 

\section{Experimental method}
\begin{figure}[btp]
\begin{center}
\leavevmode
\includegraphics[width=0.7\linewidth]{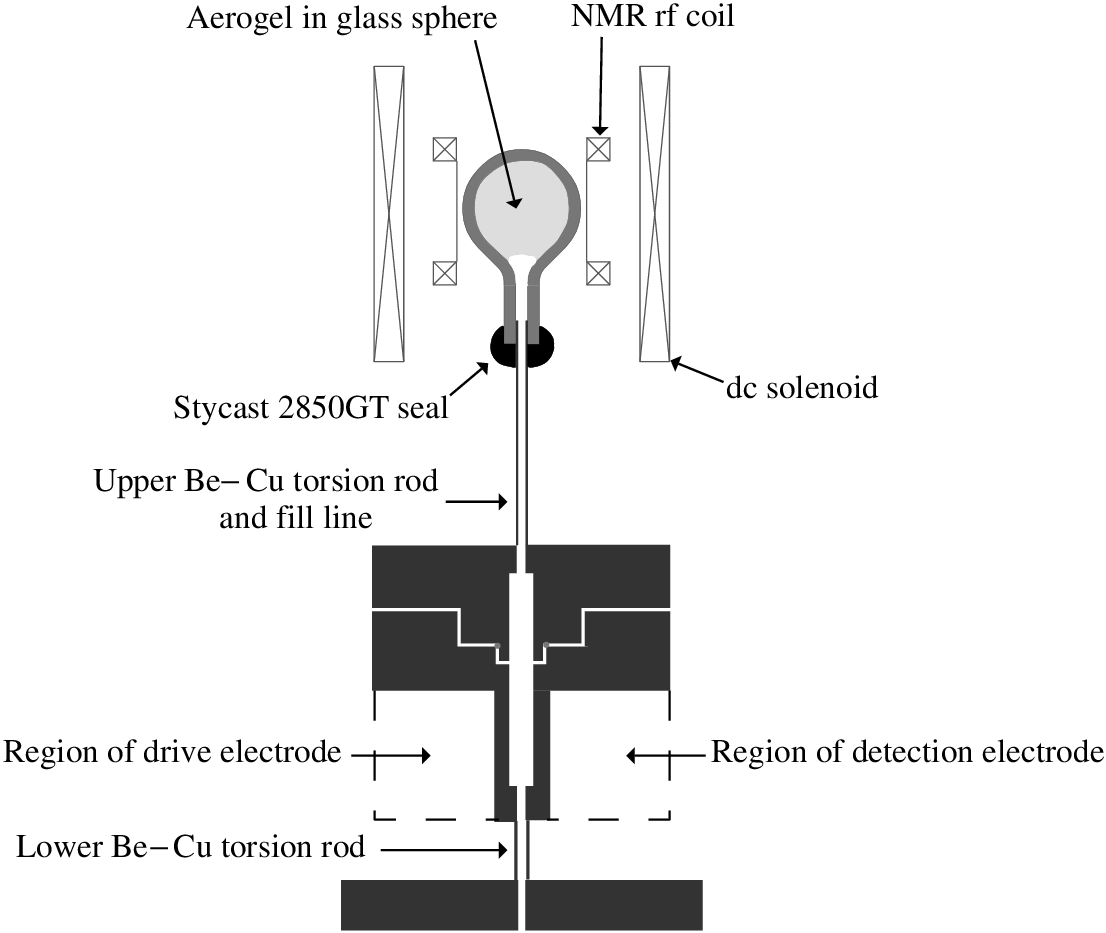}
\bigskip
\caption{Schematic diagram of the experimental setup. The large
moment of inertia of the demountable Be-Cu cylinder between the upper
and lower torsion rods together with the lower torsion rod provides a
low pass filter preventing high frequency vibrational noise reaching
the torsional oscillator. The drive and detection electrodes are
situated underneath the Be-Cu cylinder in the regions shown, thus
avoiding the need to attach electrodes to the glass sphere and also
allowing the use of electrodes of larger area.}
\label{Fig1}
\end{center}
\end{figure}

The aerogel was grown within a spherical glass container with outer
and inner diameters of order $10\,{\rm mm}$ and $8\,{\rm mm}$
respectively as shown in Fig.\,\ref{Fig1}. Two samples have been
studied with nominal densities of 1\% and 2\% of that of solid glass.
The $^3$He entered the cell via a glass stem with an internal diameter
of about $1\,{\rm mm}$. In the case of the 1\% sample the spherical
envelope was completely filled with aerogel but for the 2\% sample a
small region of bulk liquid was present near the stem as indicated in
Fig.\,\ref{Fig1}. The glass stem was glued to a berylium copper
capillary with stycast 2850 GT epoxy\cite{Emerson}; the capillary
acted as the fill line and also provided the torsion rod for torsional
oscillations of the cell. The torsional oscillations were excited and
detected electrostatically using electrodes mounted on the low pass
vibration filter shown in Fig.\,\ref{Fig1}. The vibration frequencies
for the 1\% and 2\% samples were of order $850\,{\rm Hz}$ and
$970\,{\rm Hz}$ respectively. A two-phase lockin amplifier was used to
measure the output from the detection electrode; from the in-phase and
quadrature signals the resonant frequency and bandwidth of the
oscillator were determined. Computer control of the driving frequency
and voltage ensured that the oscillator was driven close to its
resonant frequency and at a user-specified amplitude\cite{Gould89}.

We were also able to simultaneously make transverse NMR measurements
on the $^3$He at frequencies up to about $200\,{\rm kHz}$; a full
account of the NMR experiments will be given later. Temperature was
measured using an LCMN thermometer mounted in a separate tower. The
LCMN was calibrated against the transition temperature of bulk
$^3$He\cite{Greywall86}; the superfluid transition of the $^3$He in the
thermometer was identified from the change in warming rate of the LCMN.
As we were able to detect the bulk superfluid transition in both the
thermometer and the glass sphere we were able to monitor and correct
for the small temperature differences between thermometer and aerogel.
\\

\section{Superfluid transition temperature}
\begin{figure}[btp]
\begin{center}
\leavevmode
\includegraphics[width=0.8\linewidth]{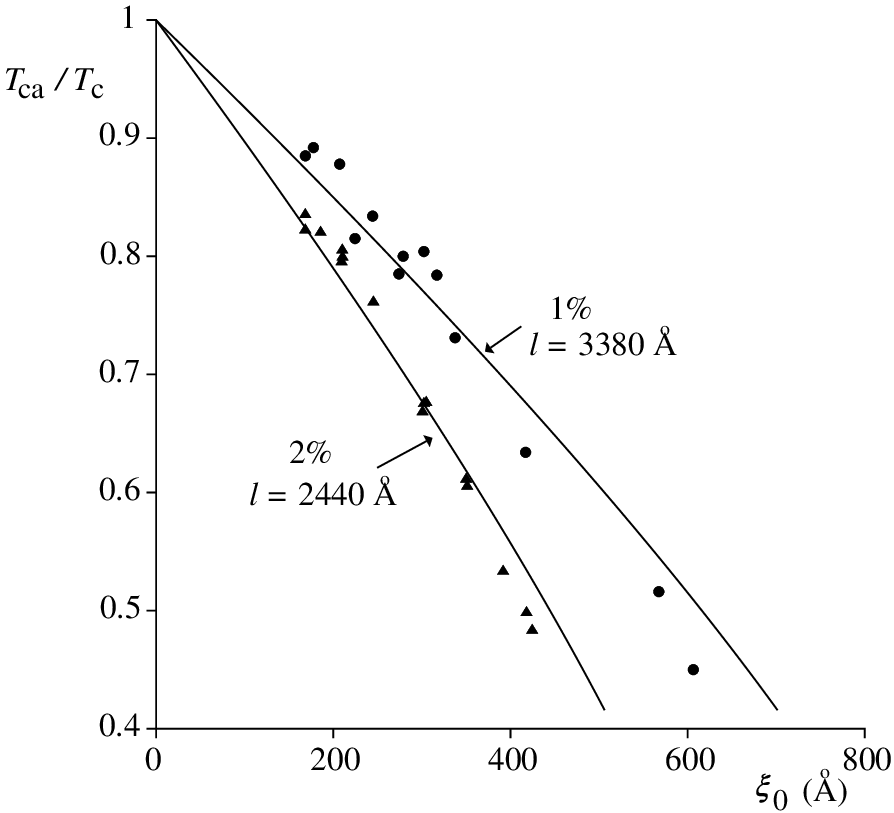}
\bigskip
\caption{Superfluid transition temperature for $^3$He in aerogel at
various pressures. The circles and triangles are for 1\% and 2\%
aerogel respectively. The theoretical curves are fits of the data to 
Eq.\,(\protect{\ref{2.2}}) using values of the mean free path $l$ of
$3380\,\AA$ and $2440\,\AA$ for the 1\% and 2\% aerogel respectively.}
\label{Fig2}
\end{center}
\end{figure}
Fig.\,\ref{Fig2} shows the superfluid transition temperature $T_{\rm
ca}$ of the $^3$He in the aerogel at various pressures relative to the
superfluid transition temperature $T_{\rm c}$ of bulk $^3$He. The
values of $T_{\rm ca}$ are plotted as functions of the bulk coherence
length which we define as
\begin{equation}
\xi_0={{\hbar v_{\rm F}}\over{2\pi k_{\rm B}T_{\rm c}}}.
\label{2.1}
\end{equation}
For the 2\% aerogel the values of $T_{\rm ca}$ were obtained from the
torsional oscillator frequency measurements as described in Sec. 4.
For the 1\% aerogel the values of $T_{\rm ca}$ were those below which
a shift in NMR frequency from the Larmor value was observed; as
explained in Sec. 5. it was not possible to use the torsional
oscillator frequency to identify the transition reliably at all
pressures.

To the extent that the aerogel behaves like a dilute impurity which
subjects the $^3$He superfluid to isotropic pair breaking scattering
the reduction in transition temperature should be given by the
following implicit equation\cite{Abrikosov60,Thuneberg?}
\begin{equation}
\ln \left({{T_{\rm ca}}\over{T_{\rm
c}}}\right)=\psi\left({1\over2}\right)
-\psi\left({1\over2}+{{\xi_0T_{\rm c}}\over{lT_{\rm ca}}}\right),
\label{2.2}
\end{equation}
where $\psi(x)$ is the digamma function and $l$ is the mean free path
associated with the scattering. We have adjusted $l$ to
obtain the best possible fits to the experimental data; the values
required were $3380\,\AA$ and $2440\,\AA$ for the 1\% and 2\% aerogel
respectively. As has also been observed in other
experiments\cite{Thuneberg?,Porto96}, there is a small but systematic
difference between the observed pressure dependence and that predicted
by Eq.\,(\ref{2.2}) although this could be evidence for a small
pressure dependence of $l$. The values of $l$ are close to those
obtained in a naive calculation for collisions of $^3$He atoms with
silica strands of diameter $3\,{\rm nm}$\cite{Thuneberg?}. Such a
calculation predicts that $l$ should be inversely proportional to the
aerogel density; that our fitted values of $l$ differ by less than a
factor of two probably indicates that the real densities of our
aerogel samples differ somewhat from their nominal values. We note
that other measurements of $T_{\rm ca}$ for aerogels with nominal
density 2\% of solid show considerable variation. Our values of
$T_{\rm ca}$ lie between those of Ref.\,\cite{Porto95} and those of
Refs. \cite{Halperin95} and \cite{Matsumoto97}.

One important deduction that can be made from the success of
Eq.\,(\ref{2.2}) in describing the depression of the superfluid
transition is that this indicates that the Cooper pairing within the
aerogel is of the same type as in bulk, namely p-wave.

\section{Measurements for 2\% aerogel}

\begin{figure}[btp]
\begin{center}
\leavevmode
\includegraphics[width=0.7\linewidth]{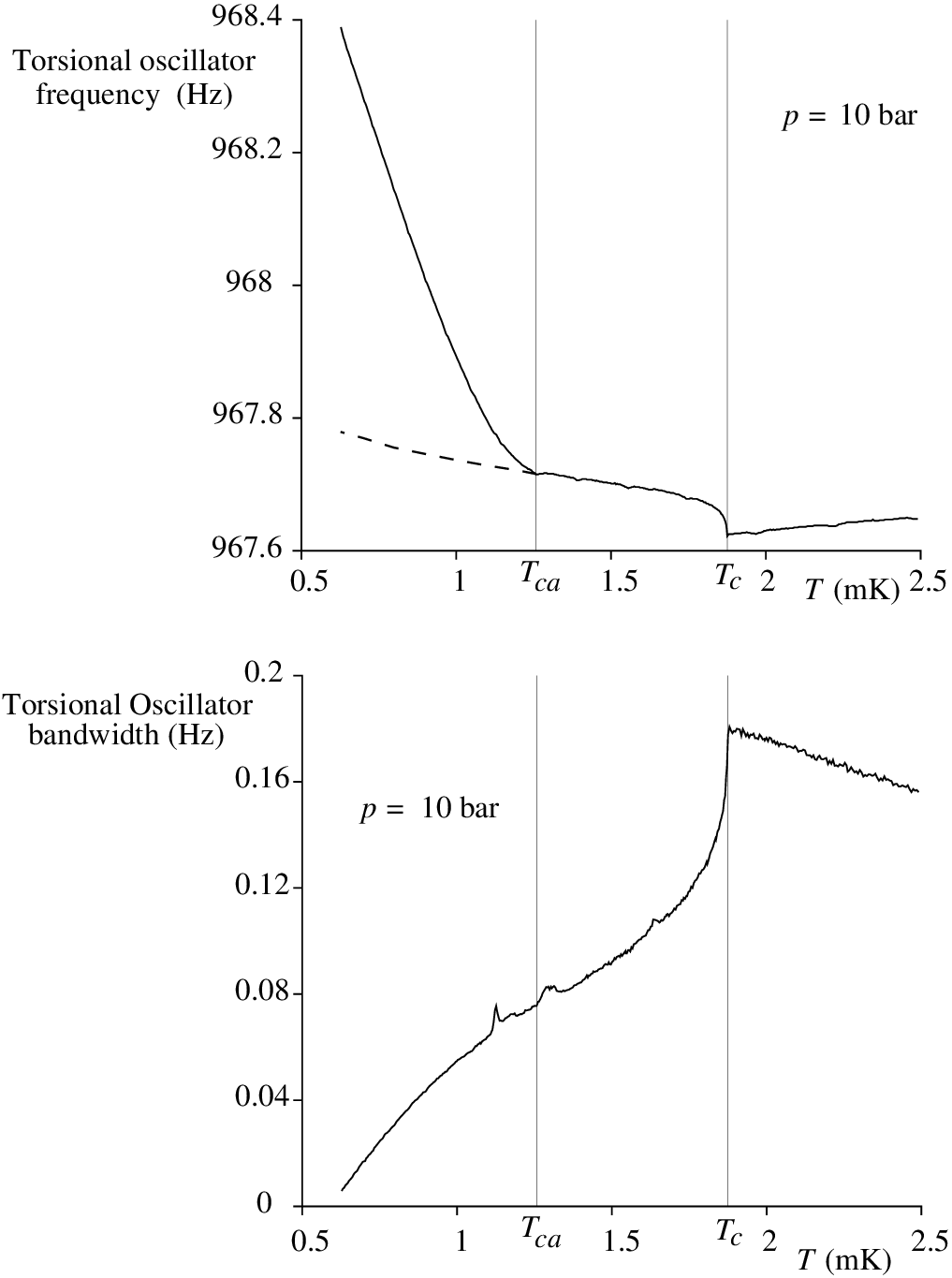}
\bigskip
\caption{Resonant frequency and bandwidth of the torsional
oscillator as a function of temperature at $10\,{\rm bar}$
pressure for 2\% aerogel.}
\label{Fig3}
\end{center}
\end{figure}
\begin{figure}[btp]
\begin{center}
\leavevmode
\includegraphics[width=0.7\linewidth]{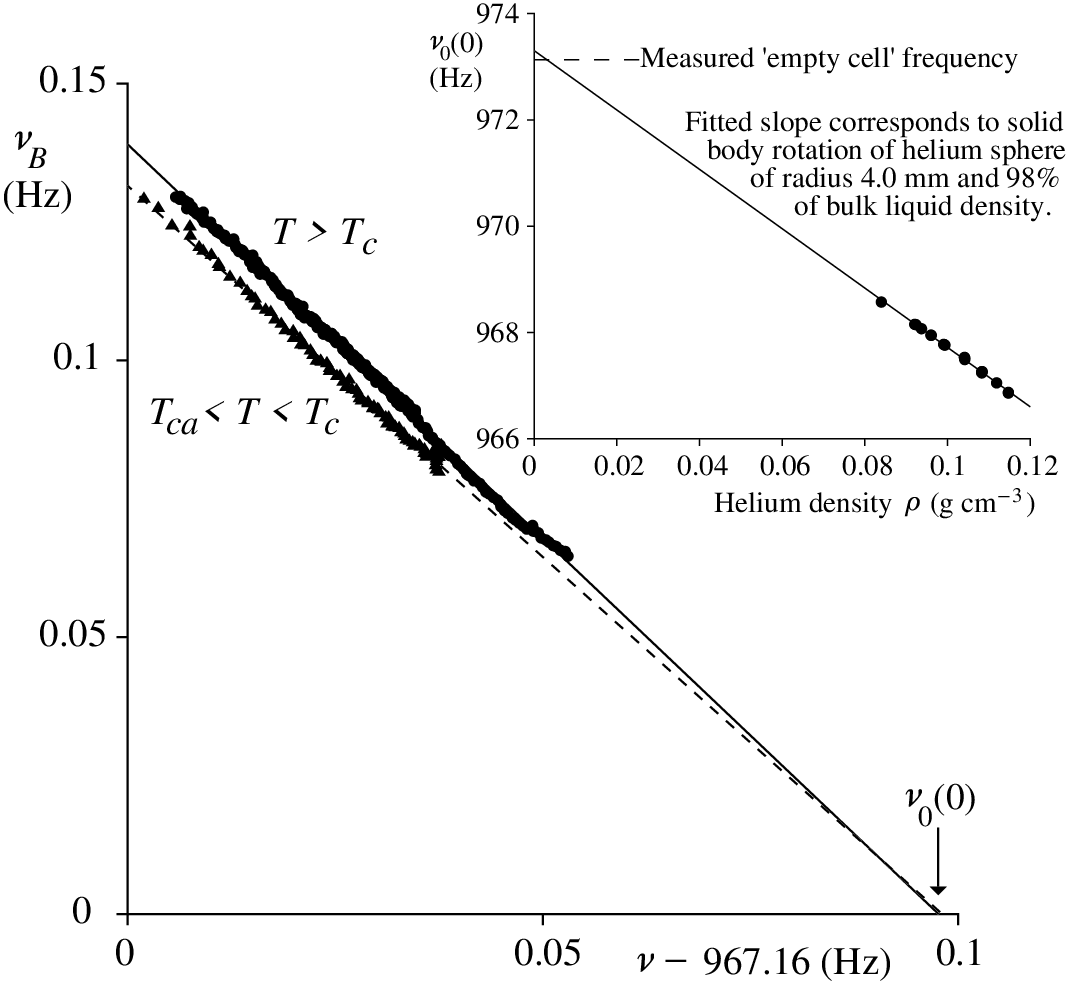}
\bigskip
\caption{Bandwidth versus resonant frequency of the torsional
oscillator for $T>T_{\rm c}$ (filled circles) and for $T>T_{\rm ca}$
(filled triangles). Extrapolation of the linear relation to zero
bandwidth as shown enables the frequency $\nu_0(0)$ to be determined.
Values of $\nu_0(T)$ for $T<T_{\rm ca}$ were obtained by using the
measured bandwidth at $T$ and the extrapolation of the $T>T_{\rm ca}$
data. The inset (top right) shows values of $\nu_0(0)$ as a function of
$\rho$; the straight line fit has an intercept which agrees with the
empty cell frequency within experimental error.}
\label{Fig4}
\end{center}
\end{figure}
Fig.\,\ref{Fig3} shows measurements of bandwidth and resonant
frequency of the torsional oscillator for 2\% aerogel filled with
$^3$He at $10\,{\rm bar}$ pressure; the temperatures, $T_{\rm c}$ and
$T_{\rm ca}$, of the superfluid transitions of the small region of
bulk $^3$He and of the $^3$He in the aerogel are indicated. Above
$T_{\rm ca}$ the temperature dependence is associated entirely with
the small region of bulk liquid; the aerogel and $^3$He within it
behave as though rigidly locked to the motion of the oscillator. The
increase in resonant frequency below $T_{\rm ca}$ signifies the
decoupling of the superfluid fraction within the aerogel from the
motion of the oscillator. The superfluidity of the $^3$He within the
aerogel does not produce any additional dissipation; apart from the
small peak in bandwidth just below $T_{\rm ca}$ and scarcely visible
increments in dissipation even closer to $T_{\rm ca}$, which are
discussed further below, the dissipation is associated entirely with
the bulk liquid region. There is no evidence that the oscillator
frequency and bandwidth depend significantly on magnetic fields up to
51 Gauss, the value used in most of our NMR measurements.

To determine the reduced superfluid density $\rho_{\rm s}/\rho$ within
the aerogel we use
\begin{equation}
{\rho_{\rm s}\over\rho}={{\nu(T)-\nu_{\rm 0}(T)}\over
{\nu_{\rm e}-\nu_{\rm 0}(0)}},
\label{3.1}
\end{equation}
where $\nu(T)$ is the measured resonant frequency at temperature $T$ as
indicated by the solid line on Fig.\,\ref{Fig3}, $\nu_{\rm 0}(T)$ is
the frequency indicated by the dashed line that would be expected at
temperature $T$ in the absence of superfluidity in the aerogel and
$\nu_{\rm e}$ the measured frequency of the cell prior to filling with
$^3$He; $\nu_{\rm 0}(0)$ is the frequency to be expected for the case
where the $^3$He within the aerogel is rigidly coupled to the motion of
the oscillator but the bulk $^3$He is completely decoupled. The value
of $\nu_{\rm 0}(0)$ could be obtained from both the normal state data
at sufficiently high temperatures and the data just above $T_{\rm ca}$;
in both these cases the viscous penetration depth
$\delta=\sqrt{2\eta/\rho_{\rm n}\omega}$ is small compared to the size
of the bulk liquid region leading to linear relations between
bandwidth and resonant frequency as shown in Fig.\,\ref{Fig4}.
Extrapolation of these relationships to zero bandwidth
($\equiv\delta\rightarrow0$) allows $\nu_{\rm 0}(0)$ to be determined.
At all pressures the values obtained from the two extrapolations agreed
within an experimental uncertainty of about $10\,{\rm mHz}$. To obtain
the value of $\nu_{\rm 0}(T)$ for $T<T_{\rm ca}$ the extrapolation of
the bandwidth vs frequency for the $T>T_{\rm ca}$ data was used as
shown by the dashed line on Fig.\,\ref{Fig4}. Since the dissipation was
unaffected by the superfluid transition within the aerogel the measured
bandwidth at the temperature concerned allowed us to obtain a value for
the `unshifted' resonant frequency. This method ignores possible
departures from hydrodynamic behaviour within the bulk region at low
temperatures; since the variation of $\nu_{\rm 0}(T)$ with temperature
is very weak (see the dashed line on Fig.\,\ref{Fig3}), this is
unlikely to lead to significant error. The values obtained for
$\nu_{\rm 0}(0)$ at different pressures are shown in the inset on
Fig.\,\ref{Fig4} as functions of the density $\rho$ of bulk $^3$He at
the pressure concerned; the straight line through the data corresponds
to that expected for rigid torsional oscillations of a sphere of
diameter $4\,{\rm mm}$ and density $0.98\rho$. The extrapolation of the
line at $\rho=0$ to a value close to the measured empty cell frequency
confirms our belief that the aerogel in the 2\% sample moves rigidly
with the glass envelope.

\begin{figure}[bt]
\begin{center}
\leavevmode
\includegraphics[width=0.6\linewidth]{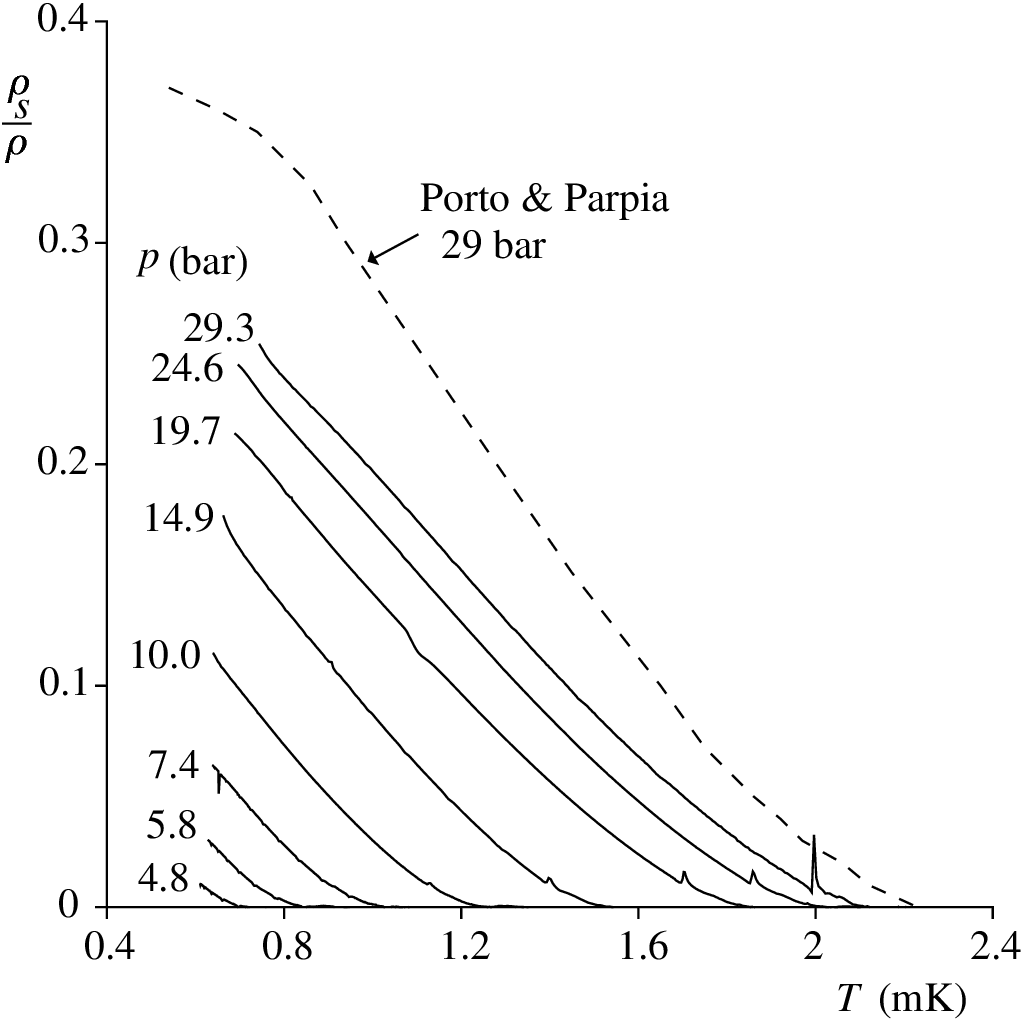}
\bigskip
\caption{Temperature dependence of $\rho_{\rm s}/\rho$ for 2\% aerogel
at the pressures indicated. The data of Porto and Parpia at $29\,{\rm
bar}$ are shown by the dashed line.}
\label{Fig5}
\end{center}
\end{figure}

Values of $\rho_{\rm s}/\rho$ at different pressures obtained using
Eq.\,(\ref{3.1}) are shown in Fig.\,\ref{Fig5}. The behaviour is
qualitatively similar to that observed by Porto and
Parpia\cite{Porto95,Porto96} also for aerogel of nominal 2\% of solid
density; the data from Ref.\,\cite{Porto96} at $29\,{\rm bar}$ are
also shown on Fig.\,\ref{Fig5}. The temperature dependence of
$\rho_{\rm s}/\rho$ is very different to that of bulk $^3$He. It rises
more slowly as $T$ decreases through $T_{\rm ca}$ and approaches a
value substantially less than unity at the lowest temperatures; this
latter feature is more apparent in the data of Ref.\,\cite{Porto96}
than in our data. Our data share with that of Ref. \cite{Porto96} the
character that close to $T_{\rm ca}$ the values of $\rho_{\rm s}/\rho$
at different pressures can be superimposed to a good approximation by
a translation along the temperature axis. For $T/T_{\rm ca}>0.75$,
$\rho_{\rm s}/\rho$ is well fitted by a temperature dependence
\begin{equation}
{\rho_{\rm s}\over\rho}=A\left(1-{T\over{T_{\rm ca}}}\right)^b.
\label{3.2}
\end{equation}

\begin{figure}[ht]
\begin{center}
\leavevmode
\includegraphics[width=0.6915\linewidth]{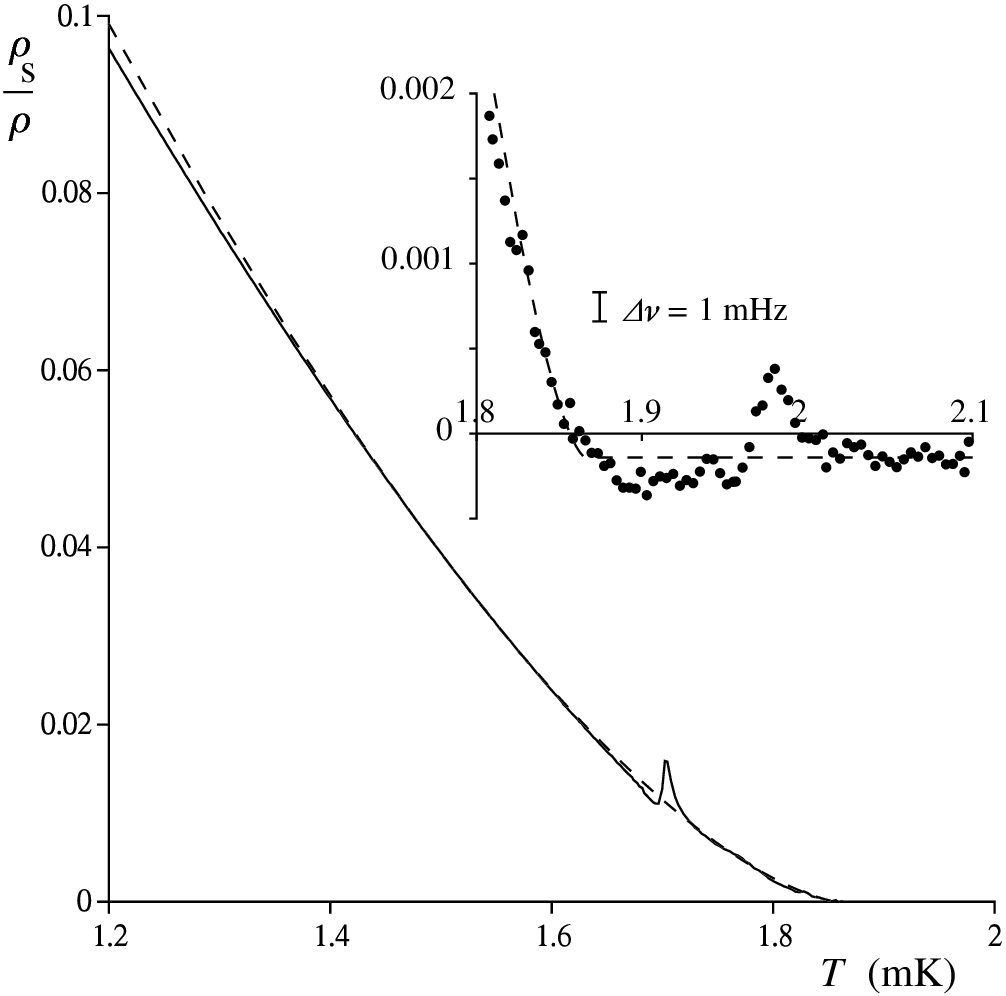}
\bigskip
\caption{Fit of $\rho_{\rm s}/\rho$ to Eq.\,(\protect{\ref{3.2}}) at
$19.7\,{\rm bar}$. The inset shows the behaviour close to $T_{\rm
ca}$; the apparent value of a small negative $\rho_{\rm s}/\rho$ at
$T>T_{\rm ca}$ indicates a small error in the  value of $\nu_{\rm
0}(T)$
used in the determination of $\rho_{\rm s}/\rho$ through
Eq.\,(\protect{\ref{3.1}}). In both figures the fitted curve is shown
with a dashed line.}
\label{Fig6}
\end{center}
\end{figure}

\begin{figure}[btp]
\begin{center}
\leavevmode
\includegraphics[width=0.692\linewidth]{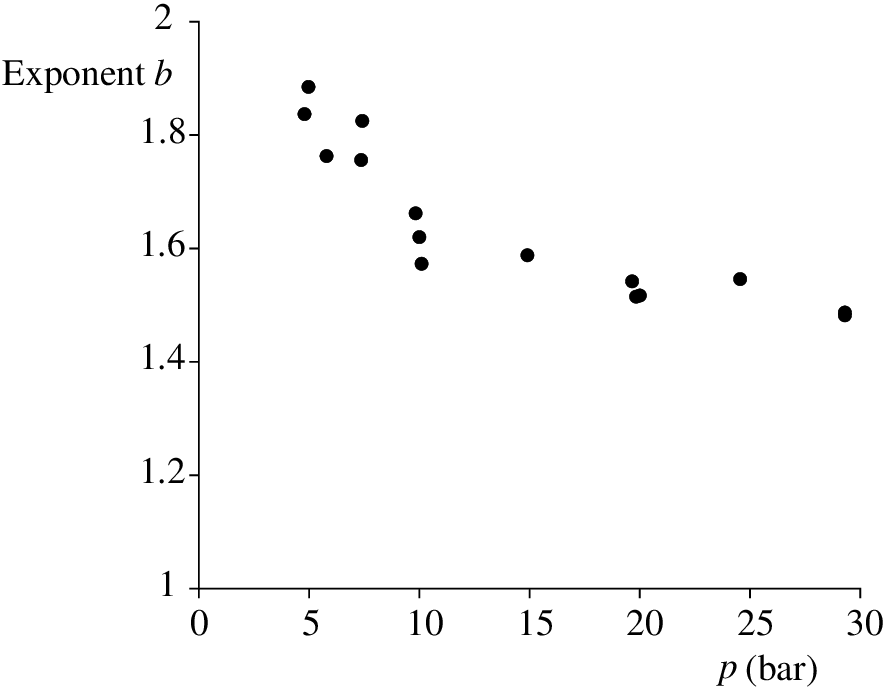}
\bigskip
\caption{Values of the exponent $b$ giving the temperature dependence
of $\rho_{\rm s}$ near $T_{\rm c}$.}
\label{Fig7}
\end{center}
\end{figure}

\begin{figure}[btp]
\begin{center}
\leavevmode
\includegraphics[width=0.7\linewidth]{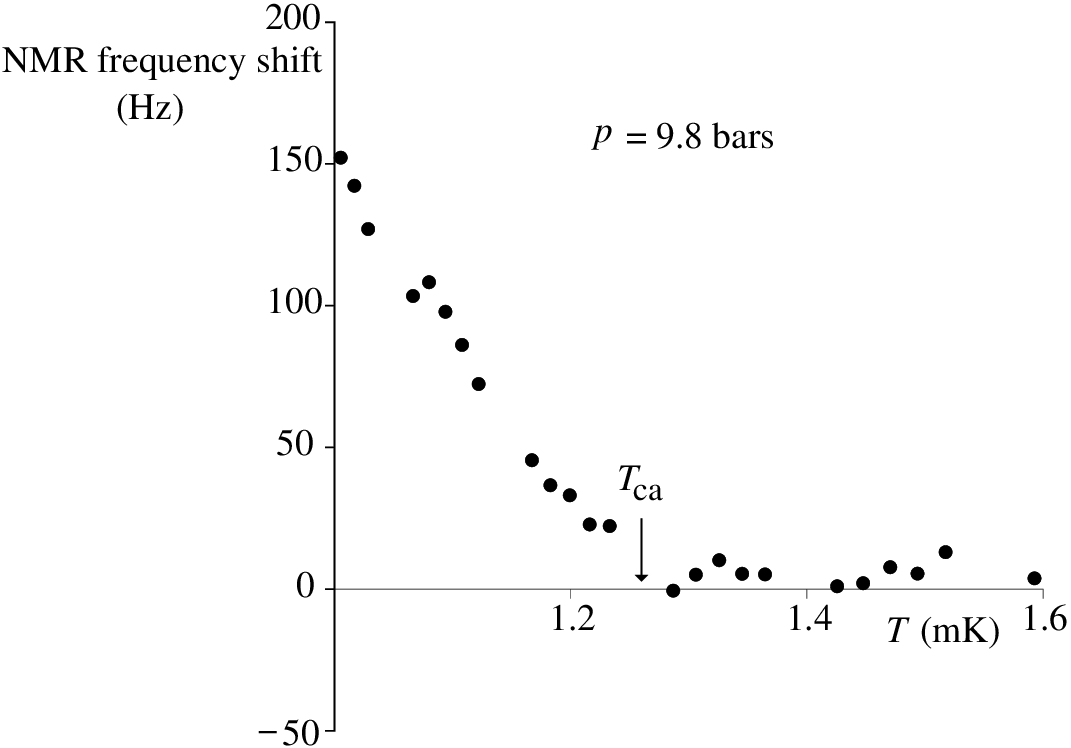}
\bigskip
\caption{NMR frequency shift as a function of temperature. The value
of $T_{\rm ca}$ indicated was obtained from the torsional oscillator
measurements.}
\label{Fig8}
\end{center}
\end{figure}

An example of such a fit is shown in Fig.\,\ref{Fig6}. The values of
the exponent $b$ at different pressures are given in Fig.\,\ref{Fig7}.
Fits to the bulk superfluid density over the same range of reduced
temperatures do not fit Eq.\,(\ref{3.2}) well near $T_{\rm c}$ (where
the correct limiting behaviour is $\rho_{\rm s}/\rho\propto(1-T/T_{\rm
c})$) and produce significantly smaller values of $b\sim1.25-1.30$. We
note that Porto and Parpia obtained somewhat smaller values of $b$
than ours from logarithmic fits ($\ln(\rho_{\rm s}/\rho)$ vs
$\ln(1-T/T_{\rm ca})$); also their values of $b$ increased with
increasing pressure. We used our fits to Eq.\,(\ref{3.2}) as an
objective way of determining the values of $T_{\rm ca}$ for the 2\%
aerogel; this proved to be a more accurate way than from the NMR
frequency shift data. There is no evidence for a systematic difference
between the transition temperatures observed by the two experimental
methods as can be seen from Fig.\,\ref{Fig8} which compares NMR
frequency shift data with the value of $T_{\rm ca}$ obtained from the
torsional oscillator.

Finally in this section we discuss the peak in bandwidth which can be
seen just below $T_{\rm ca}$ on Fig.\,\ref{Fig3}. This peak was seen
at all pressures above $7\,{\rm bar}$ but was largest at high
pressures where a corresponding feature in the resonant frequency
produced the small glitches in ${\rho_{\rm s}/\rho}$ on
Fig.\,\ref{Fig5}. The occurrence of the peak just below $T_{\rm ca}$
suggests a mode crossing of the torsional oscillator frequency with an
internal mode of oscillation of the cell which has a frequency
increasing from zero as $\rho_{\rm s}$ becomes finite within the
aerogel. A second smaller peak in dissipation was observed even closer
to $T_{\rm ca}$; there was evidence also for small additional
dissipation between this second peak and $T_{\rm ca}$. These
observations suggest the existence of several internal modes such as
might be associated with sound waves in the aerogel.

\clearpage

\begin{figure}[btp]
\begin{center}
\leavevmode
\includegraphics[width=0.7\linewidth]{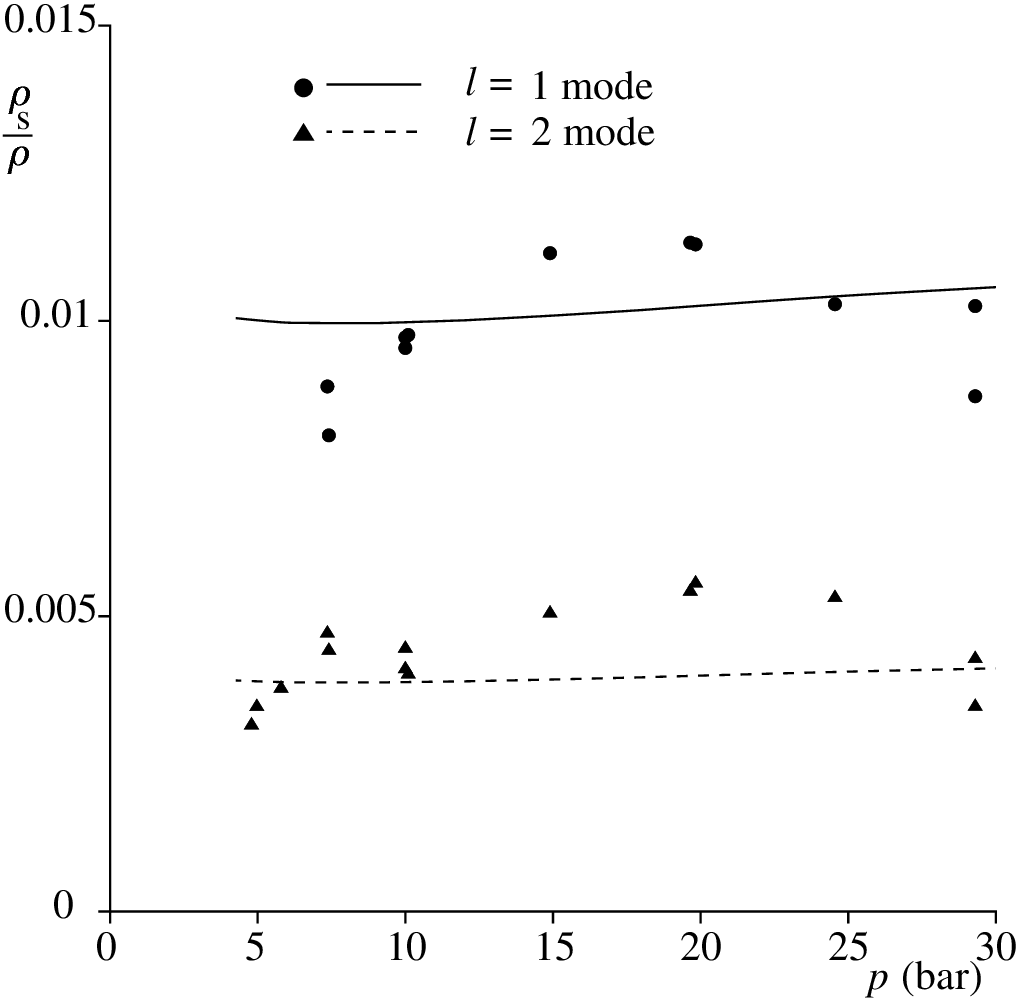}
\bigskip
\caption{Comparison between the experimental (circles and triangles)
and theoretical (continuous and dashed line) values of
$\rho_{\rm s}/\rho$ at which mode crossing of the torsional oscillator
frequency with sound modes within the sphere occur.}
\label{fig9}
\end{center}
\end{figure}

Following McKenna et al\cite{McKenna91}, we assume that the normal
fluid is rigidly locked to the aerogel. The sound speeds, $c$, are
then the roots of the equation
\begin{equation}
c^4-c^2(c_1^2+c_2^2)+c_1^2c_2^2+{{\rho_{\rm a}}\over{\rho_{\rm n}}}
(c^2-c_{\rm a}^2)(c^2-c_4^2)=0,
\label{3.3}
\end{equation}
where $c_{\rm a}$ is the speed of sound in the aerogel in the
absence of the $^3$He and $c_1$, $c_2$ and $c_4$ are the speeds of
first, second and fourth sound respectively. In $^3$He, $c_2$ is very
small and can be set to zero in Eq.\,(\ref{3.3}). One of the roots of
Eq.\,(\ref{3.3}) goes to zero as $\rho_{\rm s}\rightarrow0$ and the
sound modes in a sphere of radius $a$ associated with this root are
calculated in appendix A. The two lowest modes occur at angular
frequencies $2.082c/a$ and $3.342c/a$ and lead to mode crossings with
the torsional oscillator frequency at values of $\rho_{\rm s}/\rho$
given by the theoretical lines on Fig.\,\ref{fig9}.
The fit with the experimental values was obtained by adjusting $c_{\rm
a}$; the value used is $174\,{\rm m\,s}^{-1}$. McKenna et al quote a
value of $c_{\rm a}\sim 100\,{\rm m\,s}^{-1}$ for aerogel of 5\% of
solid density. Our value seems higher than might be expected for 2\%
aerogel since $c_{\rm a}$ is expected to decrease as the density of
aerogel decreases\cite{Gross97}; it is possible that there will be
variation between different samples of the same nominal density and
also that $c_{\rm a}$ may vary with temperature. It is clear from
inspection of Fig.\,\ref{fig9} that there is a systematic difference
in pressure dependence between the theoretical curves and experimental
points. We have no explanation of this although we note that the
experimental values at low pressures are not well determined because
the peaks were very small and somewhat broader. We note that the
existence of coupling between the torsional oscillations and the sound
modes indicates that our experiment does not have perfect rotational
symmetry about a vertical axis.

We consider also the possibility that the intersecting mode is a
Helmholtz resonance with pressure oscillations within the glass sphere
producing oscillatory flow through the fill line. If our spherical
container were filled only with $^3$He then the geometry of our fill
line is such that the Helmholtz frequency would always be more than a
factor of two less than that of the torsional oscillator even when the
superfluid fraction is 100\%. The introduction of aerogel will decrease
the frequency and hence the intersecting mode is not likely to be a
Helmholtz resonance. There is a possibility that the sound modes
discussed above could be modified by flow through the fill line. In
appendix A we discuss a simple model which takes both the flow through
the fill line and the spatial variation of pressure within the aerogel
into account. The conclusion is that the fill line impedance is
sufficiently high that flow through the fill line is unlikely to have
had a significant effect on the sound mode frequencies.

\section{Measurements for 1\% aerogel}
\begin{figure}[btp]
\begin{center}
\leavevmode
\includegraphics[width=0.8\linewidth]{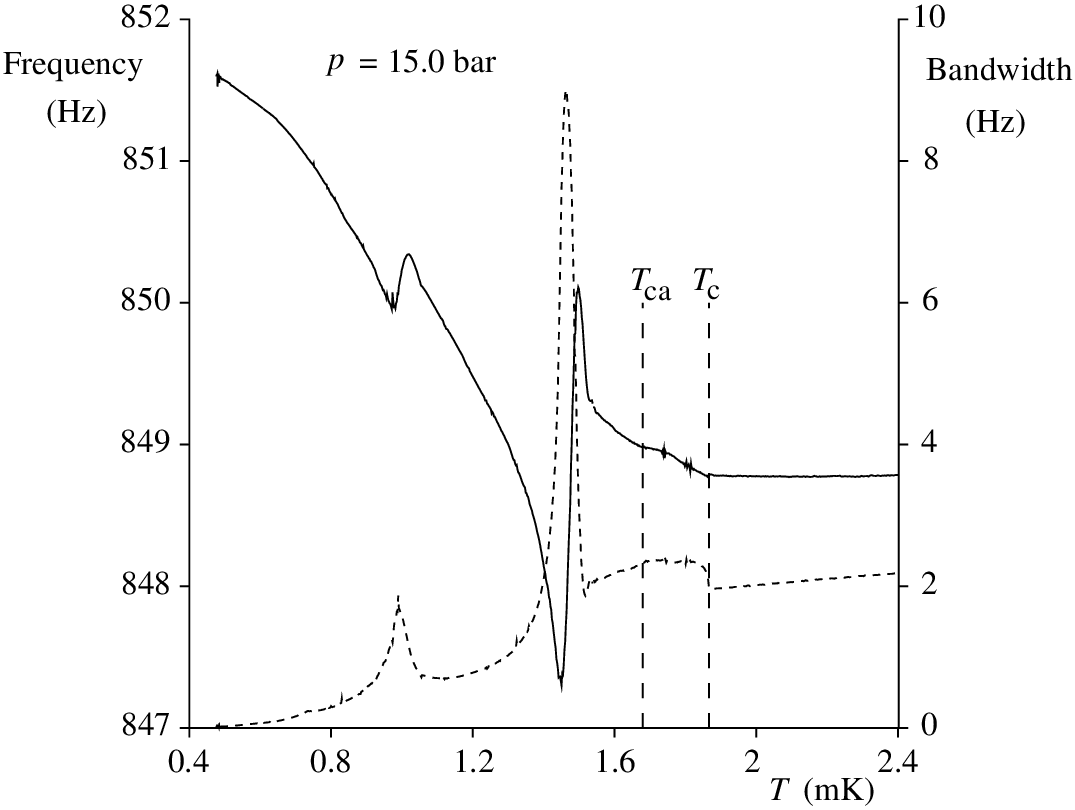}
\bigskip
\caption{Resonant frequency (continuous line) and bandwidth (dashed
line) of the torsional oscillator at $15\,{\rm bar}$ pressure for the
1\% aerogel.}
\label{fig10}
\end{center}
\end{figure}

Fig.\,\ref{fig10} shows the bandwidth and the resonant frequency as
functions of temperature at $15.0\,{\rm bar}$ pressure for the 1\%
aerogel.
The bulk superfluid transition is clearly seen in both the amplitude
and bandwidth despite the absence of a macroscopic bulk superfluid
region inside the sphere. The superfluid transition in the aerogel as
determined from the NMR frequency shift is indicated and corresponds
to an upturn in the resonant frequency with decreasing temperature
just as for the 2\% aerogel (Fig.\,\ref{Fig3}). However, just below
$T_{\rm ca}$, the behaviour of the torsional oscillator is dominated
by coupling to another resonant mode. We do not believe that this
parasitic mode is a sound mode like those observed for the 2\%
aerogel. Indeed it may not be associated uniquely with the onset of
superfluidity of the $^3$He in the aerogel since at low pressures the
coupling to this mode appears to be evident above $T_{\rm ca}$; it is
this coupling above $T_{\rm ca}$ which prevented identification of
$T_{\rm ca}$ from the torsional oscillator data at low pressures. The
large magnitude of the coupling between the mode and the torsional
oscillations together with the large bandwidth of the torsional
oscillator at all temperatures in comparison with the 2\% data suggest
that significant shear motions of the 1\% aerogel are being excited.
Further evidence for this is the apparent discontinuity in the
`background' frequency of the torsional oscillator associated with the
mode crossing. This possible discontinuity prevents the deduction of
reliable values of superfluid density for the 1\% aerogel at lower
temperatures although we are currently searching for a theoretical
description of the intersecting mode which might make this possible.

The smaller intersecting resonance which can be seen at lower
temperatures on Fig.\,\ref{fig10} is we believe due to a sound mode
like those observed for the 2\% aerogel. If we ignore the apparent
discontinuity in the background frequency of the torsional oscillator
mentioned in the previous paragraph we can estimate a value of
$\rho_{\rm s}$ at which this second resonance occurs and then by
proceeding as for the 2\% aerogel we can estimate the speed of sound
$c_{\rm a}$ in the 1\% aerogel. The value obtained is $55\,{\rm
m\,s^{-1}}$ which, in view of our neglect of a possible background
frequency discontinuity, must be regarded as an upper limit; the value
is less than that for the 2\% aerogel as would be expected.

For the 1\% aerogel we investigated the effect of the addition of small
quantities of $^4$He on the torsional oscillator. About 3\% of $^4$He
was sufficient to replace the solid He layer on the aerogel surfaces as
indicated by the absence of a Curie-Weiss contribution to the NMR
signal strength\cite{Halperin95}. The addition of $^4$He caused a
dramatic reduction in the size of the parasitic resonances and of the
oscillator bandwidth at higher temperatures; coupling to the
intersecting resonances was almost completely absent at high pressures.
We do not have any explanation for this observation. Although the
parasitic resonances were absent we were still unable to determine the
superfluid density within the aerogel because the addition of the
$^4$He also introduced a large thermal boundary resistance between the
LCMN and the $^3$He which preventing us from measuring the temperature.

\section{Conclusions}
 In this concluding section we return briefly to the questions posed in
 our introduction. Our experiments show that the superfluid transition
 of $^3$He in aerogel as indicated by the appearance of a finite
 superfluid density coincides with the onset of a shift in the NMR
 frequency. Our torsional oscillator measurements provide no evidence
 for more than one superfluid phase and do not identify the nature of
 the pairing although our observation that the measured superfluid
 density is independent of applied magnetic fields up to about $50\,{\rm
 Gauss}$ might be interpreted as implying that the superfluid phase is
 one with an isotropic superfluid density. We are hoping that the
 completion of our analysis of our NMR measurements will provide
 further information on the questions of the number of phases and the
 nature of the pairing. The fact that we measure two different
 properties on the same sample of aerogel should provide a stringent
 test for theories purporting to explain the effect of aerogel on the
 properties of the superfluid.

\begin{ack}
We have benefitted greatly from discussions with Henry Hall. We are
grateful to Norbert Mulders, Jongsoo Yoon and Moses Chan for providing
the aerogel specimens used in our experiments. This work was supported
by EPSRC through Research Grants GR/K59835 and GR/K58234, and by the
award of Research Studentships to JJK and PSW.
\end{ack}

\appendix
\section{Sound Modes and Helmholtz Resonance}

We consider first the sound modes in a spherical cavity of radius $a$
completely filled with aerogel containing $^3$He. The small value of
$c_2$ for $^3$He means that the effect of temperature gradients within
the cavity can be ignored and the equations of motion which describe
the $^3$He/aerogel combination are then\cite{McKenna91}
\begin{eqnarray}
{{\partial\rho}\over{\partial t}}&=&-{\bf\nabla}\cdot(\rho_{\rm s}
{\bf v}_{\rm s}+\rho_{\rm n}{\bf v}_{\rm n}),\label{A1}\\
{{\partial{\bf v}_{\rm s}}\over{\partial t\ }}
&=&-{1\over\rho}{\bf\nabla}p,\label{A2}\\
(\rho_{\rm a}+\rho_{\rm n}){{\partial{\bf v}_{\rm n}}
\over{\partial t\ }}&=&-{{\rho_{\rm n}}\over\rho}{\bf\nabla}p
-{\bf\nabla}p_{\rm a},\label{A3}\\
{{\partial\rho_{\rm a}}\over{\partial t\ }}&=&-{\bf\nabla}\cdot
(\rho_{\rm a}{\bf v}_{\rm n}),\label{A4}
\end{eqnarray}
where $\rho$ and $\rho_{\rm a}$ are the densities of helium and
aerogel, $p$ is the pressure acting on the helium and $p_{\rm a}$ is
the pressure acting on the aerogel due to its elastic distortion; the
motion of the aerogel and normal fluid are assumed to be locked
together.  Assuming that small variations in $p$ and $p_{\rm a}$ are
related to the corresponding densities by $\delta p=c_1^2\delta\rho$
and $\delta p_{\rm a}=c_{\rm a}^2\delta\rho_{\rm a}$, we obtain from
Eqs.\,(\ref{A1}) to (\ref{A4}) the following equation for small
harmonic departures, $\delta\rho=\rho^\prime\exp(i\omega t)$, of
$\rho$ from equilibrium
\begin{equation}
\nabla^4\rho^\prime+\omega^2\left({1\over{c_4^2}}
+{1\over{c_{\rm a}^2}}+{{\rho_{\rm n}c_1^2}\over
{\rho_{\rm a}c_4^2c_{\rm a}^2}}\right)\nabla^2\rho^\prime
+\omega^4{{(\rho_{\rm a}+\rho_{\rm n})}\over
{\rho_{\rm a}c_4^2c_{\rm a}^2}}\rho^\prime=0,
\label{A5}
\end{equation}
where $c_4^2=c_1^2\rho_{\rm s}/\rho$.

The solutions of Eq.\,(\ref{A5}) appropriate to our spherical geometry
are of the form
\begin{equation}
\rho^\prime=\left(Aj_l(k_1r)+Bj_l(k_2r)\right)Y_{lm}(\theta,\phi),
\label{A6}
\end{equation}
where $A$ and $B$ are constants of integration, $j_l(z)$ is a
spherical Bessel function, $Y_{lm}(\theta,\phi)$ is a spherical
harmonic and we are using the notation of Abramowitz and
Stegun\cite{Abramowitz65}; the spherical Bessel functions $y_l(k_1r)$
and $y_l(k_2r)$ can be excluded from the solution because they diverge
as $r\rightarrow0$. The wave numbers $k_1$ and $k_2$ are the roots of
\begin{equation}
k^4-\omega^2k^2\left({1\over{c_4^2}}+{1\over{c_{\rm a}^2}}
+{{\rho_{\rm n}c_1^2}\over{c\rho_{\rm a}c_4^2c_{\rm a}^2}}\right)
+\omega^4{{(\rho_{\rm a}+\rho_{\rm n})}\over
{\rho_{\rm a}c_4^2c_{\rm a}^2}}=0,
\label{A7}
\end{equation}
and are thus the wavenumbers associated with the sound speeds which
satisfy Eq.\,(\ref{3.3}) in the limit $c_2\rightarrow0$. The boundary
conditions to be applied to the solution are that
\begin{equation}
{\bf v}_{\rm n}\cdot\hat{\bf r}={\bf v}_{\rm s}\cdot\hat{\bf r}
=0\quad{\rm at}\quad r=a.
\label{A8}
\end{equation}
The radial components of the ${\bf v}_{\rm n}$ and ${\bf v}_{\rm s}$
can be found for solution (\ref{A6}) by using Eqs.\,(\ref{A1}) to
(\ref{A4}). Fortunately the boundary conditions produce no mixing of
the terms of different wavenumber
and the modes occur at wavenumbers $k$ ($=k_1$ or $k_2$)
which satisfy
\begin{equation}
\left[{{{\rm d}\ }\over{{\rm d}r}}j_l(kr)\right]_{r=a}=
j_{l-1}(ka)-{{l+1}\over{ka}}j_l(ka)=0.
\label{A9}
\end{equation}
The roots are thus characterised by two numbers $n$ and $l$, where $n$
specifies the number of radial nodes (excluding $r=0$) in
the variation of $p$. The values of
$ka$ ($n,l$) for the 7 lowest modes are 2.08157 (0,1), 3.34210 (0,2),
4.49341 (1,0), 4.51411 (0,3), 5.64670 (0,4), 5.94037 (1,1), 6.75645
(0,5). The two lowest modes with $n,l=0,1$ and $0,2$ are the modes
plotted in Fig.\,\ref{fig9}.

We now consider the possibility that the modes may be affected by flow
through the fill line. Since a spherical cavity on the end of
cylindrical fill line is difficult to calculate we consider instead a
simplified model: a spherical cavity with a fill line entering at the
centre. We take the end of the fill line to be surrounded by a small
spherical region of bulk liquid of radius $b$. Only modes of spherical
symmetry will have a pressure variation at the centre of the sphere.
For these modes the general solution of Eq.\,(\ref{A5}) is
\begin{equation}
\rho^\prime=\left(A_1j_0(k_1r)+B_1y_0(k_1r)+A_2j_0(k_2r)
+B_2y_0(k_2r)\right).
\label{A10}
\end{equation}
Because of the small region of bulk helium and the fill line at the
centre of the sphere it is no longer possible to ignore the spherical
Bessel function $y_0(z)=-\cos(z)/z$ which diverges as $z\rightarrow0$.
The boundary conditions to be applied at $r=a$ are Eqs.\,(\ref{A8}) as
before and at the boundary of the bulk liquid region, $r=b$, we
require that the variations of $p_{\rm a}$ should vanish and that the
ratio of pressure variation $\delta p=p^\prime\exp(i\omega t)$ to mass
flow {\it out} of the aerogel ${\dot M}=-4\pi b^2(\rho_{\rm s}v_{\rm
s}+\rho_{\rm n}v_{\rm n})$ should be appropriate to the geometry of
our fill line
\begin{equation}
{{\delta p(b)}\over{\dot M}}
={{i\omega\rho L}\over{\rho_{\rm sb}\sigma}},
\label{A11}
\end{equation}
where $L/\sigma$ is the ratio of fill line length to cross sectional
area and $\rho_{\rm sb}$ is the superfluid density in the fill line.
Eq.\,(\ref{A11}) follows from the equation of motion
\begin{equation}
{{\partial v_{\rm s}}\over{\partial t}}=i\omega v_{\rm s}
={{\delta p(b)}\over{L\rho}},
\label{A12}
\end{equation}
for the helium within the fill line. Note that we ignore the variation
of pressure inside the sphere of radius $b$.

Applying the boundary conditions to the solution given by
Eq.\,(\ref{A10}) leads to a rather complicated condition for
determining the frequency of the spherically symmetric modes. The
terms in Eq.\,(\ref{A10}) of wavenumbers $k_1$ and $k_2$ are no longer
decoupled. We do not discuss the details here but report the general
conclusion that the impedance of our fill line ($L/\sigma$) is
sufficiently high that the modes under consideration are essentially
sound modes with little flow through the fill line.

\end{document}